# High SNR 3D Imaging from Millimeter-scale Thick Tissues to Cellular Dynamics via Structured Illumination Microscopy

Mengrui Wang[1], Manming Shu[1], Jiajing Yan[1], Chang Liu[1], Xiangda Fu[1], Jingxiang Zhang[1], Yuchen Lin[1], Hu Zhao[2], Yuwei Huang[3], Dingbang Ma[4,5], Yifan Ge[4], Huiwen Hao[6], Tianyu Zhao[1], Yansheng Liang[1], Shaowei Wang[1], Ming Lei[1,7*]

[1] MOE Key Laboratory for Nonequilibrium Synthesis and Modulation of Condensed Matter, School of Physics, Xi'an Jiaotong University, Xi'an, 710049, China

[2] Chinese Institute for Brain Research, Zhongguancun Life Science Park, Beijing, 100086, China

[3] School of Basic Medical Sciences, Translational Medicine Institute, Xi'an Jiaotong University, Xi'an, 710049, China

[4] Interdisciplinary Research Center on Biology and Chemistry, Shanghai Institute of Organic Chemistry, Chinese Academy of Sciences, Shanghai, 201210, China

[5] Shanghai Key Laboratory of Aging Studies, Shanghai, 201210, China

[6] Huaxia Imaging Technology Co., Ltd., Beijing, 102600, China

[7] State Key Laboratory of Electrical Insulation and Power Equipment, Xi'an Jiaotong University. Xi'an, 710049, China

*Corresponding authors:
Ming Lei ming.lei@mail.xjtu.edu.cn
Tel: +86 18629031858

## Abstract

Three-dimensional (3D) fluorescence imaging provides a vital approach for study of biological tissues with intricate structures, and optical sectioning structured illumination microscopy (OS-SIM) stands out for its high imaging speed, low phototoxicity and high spatial resolution. However, OS-SIM faces the problem of low signal-to-noise ratio (SNR) when using traditional decoding algorithms, especially in thick tissues. Here we propose a Hilbert-transform decoding and space domain based high-low (HT-SHiLo) algorithm for noise suppression in OS-SIM. We demonstrate HT-SHiLo algorithm can significantly improve the SNR of optical sectioning images at rapid processing speed, and double the imaging depth in thick tissues. With our OS-SIM system, we achieve high quality 3D images of various biological samples including mouse brains, Drosophila clock neurons, organoids, and live cells. We anticipate that this approach will render OS-SIM a powerful technique for research of cellular organelles or thick tissues in 3D morphology.

**Introduction**

Three-dimensional (3D) fluorescence imaging has become an indispensable tool in biomedical research, offering the ability to visualize 3D biological structures such as cells, nerves, and organoids [1][2][3]. The quest for clarity and precision in observing thick tissues has led to the development of several advanced imaging techniques, each contributing uniquely to our understanding of biological processes and functions [1][4].

Among the prevalent 3D fluorescence imaging techniques, confocal laser scanning microscopy (CLSM) stands out for its ability to optically section specimens using a pinhole to eliminate out-of-focus light and thus obtaining background-free images [5][6]. However, its use is limited by photobleaching concerns and relatively slow imaging speeds, which are less than ideal for live-cell imaging. Spinning disk confocal microscopy (SDCM) enhances imaging speed by using multiple pinholes to scan the sample area simultaneously, but it results in out-of-focus light entering adjacent pinholes which compromises axial resolution[7][8]. Light sheet microscopy (LSM), on the other hand, involves illuminating the specimen with a thin sheet of light, allowing for rapid image acquisition with minimal photodamage, yet it has restrictions on the sample size and can suffer from uneven illumination due to scattering [8][10][11]. Light field microscopy (LFM) can image the entire volume simultaneously in a single camera frame, so it owns extremely high 3D imaging speed, but its inherent tradeoff between spatial and angular resolution leads to a sacrifice in spatial resolution [12][13][14].

Benefiting from the wide-field imaging approach, structured illumination microscopy (SIM) not only demonstrates high spatial resolution but also exhibits rapid imaging speed and low phototoxicity [15][16][17]. SIM was initially invented by Neil *et al.*, employing sinusoidal fringe patterns projected onto the sample surface through the objective lens [18]. Consequently, only the focal plane is modulated by the fringes, while the out-of-focus background attenuates to uniform illumination. Through root-mean-square (RMS) decoding algorithm, the in-focus information is extracted with background eliminated, thereby obtaining optical sections (OS). Structured illumination can also be replaced with speckle patterns, as in HiLo microscopy, where high-frequency information from wide-field images and low-frequency information from structured illumination images are combined to obtain optical sections [19][20]. In our previous work, we used a digital micromirror device (DMD) and LEDs to generate fringe illumination and realized high throughput 3D imaging [21]. A decoding algorithm based on Hilbert-transform (HT) reduced the image acquisition volume by 1/3 compared to the RMS operation, thus increasing the imaging speed by half [22]. With this OS-SIM system, we also realized 3D natural color imaging and morphological measurement of samples such as insects, meteorites, fossils, and minerals [23][24][25][26][27]. It should be noted that most of these measurements were made on reflective surfaces where the signal and modulation contrast is high.

For 3D fluorescence imaging of biological samples, however, OS-SIM faces the problem of low signal-to-noise ratio (SNR), especially in thick tissues with thickness over 20 μm [7][28]. Before the image becomes blurred due to scattering with increasing depth, the structured illumination has already degraded into uniform illumination, causing SIM to lose optical sectioning capability. Even in shallow regions, weak modulation depth of fringes with image acquisition noise can result in significant noise contamination in decoded optical sections [29]. Dang *et al.* proposed a Fourier domain based optical sectioning image reconstruction algorithm to suppress noise, but the improvements were modest and accompanied by a loss of high-frequency information, and the Fourier transform also reduced the processing speed [30]. Qiao *et al.* designed a channel attention generative adversarial network (caGAN) for denoising in 3D-SIM, but the training of neural network is time-consuming, and its applicability across different samples remains uncertain [31].

In this work, we propose an HT decoding and space domain based HiLo (HT-SHiLo) algorithm for noise suppression in OS-SIM. Simulation results demonstrate that HT-SHiLo algorithm significantly improves the SNR of optical sectioning images and reduces the processing time to 5.4 ms per layer at 2,048 × 2,048 pixels. We applied HT-SHiLo algorithm to the 3D fluorescence imaging of thick tissues including mouse brains, Drosophila clock neurons and organoids, finding that it not only achieved satisfactory noise suppression but also doubled the imaging depth of OS-SIM, up to 2.4 mm for tissue cleared sample. We further demonstrate its application in live-cell imaging, achieving long-term 3D recording of mitochondrial and migrasome dynamics, thanks to the low phototoxicity and fast imaging speed of OS-SIM. This method holds promise for enhancing the application of OS-SIM in studies of thick biological tissues and live-cell dynamics.

## Results

### HT-SHiLo algorithm for noise-free optical sectioning

To extract the in-focus section from the structured illumination image, the RMS operation necessitates three raw images with phase-shift sinusoidal fringes, while the HT decoding algorithm requires only two. In order to reduce the image acquisition time, we adopt the HT decoding algorithm for computation of optical sections. The two raw images can be written as

$$I_i = I_{out}(x, y) + I_{in}(x, y)\left[1 + m\sin(2\pi v x + \varphi_i)\right], \quad i = 1, 2 \qquad (1)$$

where $I_{out}$ and $I_{in}$ respectively represent the out-of-focus and the in-focus part of the sample, $m$ denotes the modulation depth, $v$ is the spatial frequency, and $\varphi_i$ is the spatial phase. Subtraction of the two raw images gives the modulated in-focus part:

$$\Delta I = I_1 - I_2 = C I_{in}(x, y)\cos(2\pi v x + \varphi_s) \qquad (2)$$

where $C = 2m\sin\dfrac{\varphi_1 - \varphi_2}{2}$, $\varphi_s = \dfrac{\varphi_1 + \varphi_2}{2}$. To solve $I_{in}(x, y)$, Hilbert transform is applied

due to its characteristic of π/2 phase-shifting:

$$\mathcal{F}\{\mathcal{H}\{f(t)\}\} = \mathcal{F}\{f(t)\}\ \mathcal{F}\left\{\frac{1}{\pi t}\right\} = -i\tilde{f}(k)\,\mathrm{sgn}(k) \tag{3}$$

$$\mathrm{sgn}(k) = \begin{cases} 1, & k>0 \\ 0, & k=0 \\ -1, & k<0 \end{cases} \tag{4}$$

where $\mathcal{F}$ and $\mathcal{H}$ respectively represent the Fourier transform and Hilbert transform operator, sgn($k$) is the signum function and $k$ represents the coordinate in frequency domain. Consequently, the Hilbert transform of $\Delta I(x, y_0)$ in $x$-direction is

$$\begin{aligned}
\mathcal{H}_x\{\Delta I(x, y_0)\} &= -i\frac{C}{2}\mathcal{F}_x^{-1}\left\{\left[\tilde{I}_{\mathrm{in}}(k_x-v, y_0)e^{i\varphi_s} + \tilde{I}_{\mathrm{in}}(k_x+v, y_0)e^{-i\varphi_s}\right]\mathrm{sgn}(k_x)\right\} \\
&\approx -i\frac{C}{2}\mathcal{F}_x^{-1}\left\{\left[\tilde{I}_{\mathrm{in}}(k_x-v, y_0)e^{i\varphi_s} - \tilde{I}_{\mathrm{in}}(k_x+v, y_0)e^{-i\varphi_s}\right]\right\} \\
&= C I_{\mathrm{in}}(x, y_0)\sin(2\pi v x + \varphi_s)
\end{aligned} \tag{5}$$

In the approximation of the second step, the negative frequency component of $\tilde{I}_{\mathrm{in}}(k_x - v, y_0)$ and the positive frequency component of $\tilde{I}_{\mathrm{in}}(k_x + v, y_0)$ were ignored, which means Equation (5) is accurate only for low-frequency information in $x$-direction ($|k_x| < v$). Nevertheless, the optical section $OS$ can be approximated through Eqs (2), (5):

$$OS = C I_{\mathrm{in}} \approx \sqrt{\Delta I^2 + \mathcal{H}_x\{\Delta I\}^2} \tag{6}$$

To achieve the best SNR for the optical section, i.e., maximizing $C$, the phase difference $\varphi_1 - \varphi_2 = \pi$ is employed in practice. In this case, the wide-field image $WF$ can be simultaneously obtained:

$$WF = (I_1 + I_2)/2 \tag{7}$$

As mentioned above, high-frequency information in $\Delta I$ undergoes a deviation after the Hilbert transform. Meanwhile, according to Eq. (6), low modulation depth will lead to low SNR of the optical section if the raw images are affected by a certain noise, as encountered in biological fluorescence imaging. Directly applying denoising methods to the optical section is difficult to achieve a desired result. However, the wide-field image is not affected by low modulation depth, providing us with an approach to denoise and reconstruct optical section with assistance of the wide-field image.

Inspired by HiLo microscopy, we propose an HT decoding and space domain based HiLo (HT-SHiLo) algorithm, which can remove the noise and restore the high-frequency information in optical sections. Figure 1 shows the flowchart of HT-SHiLo algorithm. With two raw images $I_1$, $I_2$ recorded, $OS_{HT}$ is firstly calculated through HT decoding, and the wide-field image $WF$ is also obtained. The $OS_{HT}$ and $WF$ are respectively filtered by Gaussian low pass filter ($GLPF_{\sigma=1/2\nu}$) and Gaussian high pass filter ($GHPF_{\sigma=1/2\nu}$) to acquire the low-frequency component $OS_{Lo}$ and high-frequency component $WF_{Hi}$. Since the high-frequency information in wide-field image can only come from the focal plane, it should be similar to the high-frequency information in optical section. Therefore, a noise-free optical section can be derived from weighted summation of $OS_{Lo}$ and $WF_{Hi}$:

$$OS_{HT\text{-}SHiLo} = OS_{Lo}/\eta + WF_{Hi} \qquad (8)$$

where the scaling factor $\eta$ is employed as to ensure a proportional correspondence between high-frequency and low-frequency component.

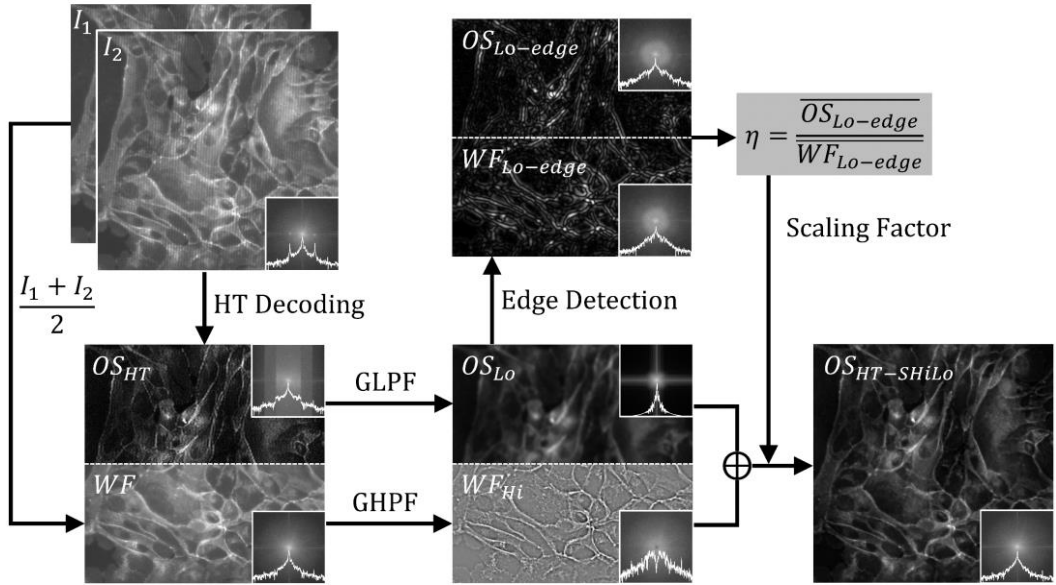

**Figure 1. Flowchart of HT-SHiLo algorithm.** HT, Hilbert transform; OS, optical sectioning image; WF, wide field image; GLPF, Gaussian low pass filter; GHPF, Gaussian high pass filter; Lo, low-frequency; Hi, high-frequency.

The standard deviation $\sigma$ (or filter radius, chosen to be $2\sigma$) of Gaussian filter determines the proportion of low-frequency component from $OS_{HT}$ to the high-frequency component from $WF$ in the final recombined optical section. As the filter radius increases, the denoising effect will improve, but the optical sectioning strength will degrade. Therefore, a trade-off should be taken to choose a proper filter radius. Here, the period of fringe on raw images $1/v$ is chosen as the filter radius for three reasons: As higher-frequency fringe will result in thinner depth of optical section [16], adjusting the filter radius accordingly ensures a matching sectioning strength; The inaccurate high-frequency ($|k_x| > v$) information in $OS_{HT}$ can be entirely replaced by $WF_{Hi}$; The residual fringes in $OS_{HT}$, whose period is the same as fringe on raw images, can be eliminated.

In the traditional HiLo microscopy, the low-pass and high-pass filtering are operated in Fourier domain, and the scaling factor is determined by ensuring a seamless transition between the low-pass and high-pass components at the cutoff frequency [19]. To avoid the time-consuming Fourier transform, HT-SHiLo algorithm performs entirely in space domain, including the calculation of the scaling factor $\eta$:

$$\eta = \frac{\overline{OS_{\text{Lo-edge}}}}{\overline{WF_{\text{Lo-edge}}}} = \frac{\overline{|GHPF_{\sigma=0.5}(OS_{\text{Lo}})|}}{\overline{|GHPF_{\sigma=0.5}(WF_{\text{Lo}})|}} \quad (9)$$

where $WF_{\text{Lo}} = WF - WF_{\text{Hi}}$. The operation $|GHPF(\ )|$ is similar to edge detection, which figures out the transition edges between $WF_{\text{Hi}}$ and $OS_{\text{Lo}}$. $WF_{\text{Lo-edge}}$ and $OS_{\text{Lo-edge}}$ are similar in distribution but differ in intensity, so the ratio of their average intensity is suitable for determination of the scaling factor.

To demonstrate the influence of modulation depth on optical sectioning, simulations were conducted on the HT decoding algorithm, HiLo algorithm and HT-SHiLo algorithm. Fringes with varying modulation depths $m$ and Gaussian white noise with a variance of 0.001, which is close to the usual situation in practice, were added to the raw image, and then optical sections were computed using the three algorithms respectively. As modulation depth decreased, the noise in HT decoded image significantly increased, while the images computed by HiLo algorithm and HT-SHiLo algorithm were less affected (Figure 2(a)). Under the lowest modulation depth $m = 0.05$, the HT decoded image was seriously eroded by noise, while the image computed by

HiLo algorithm appeared intensity bias and edge delineation (Figure 2(b)); Only the image computed by HT-SHiLo algorithm exhibited the highest similarity to the ground truth (GT) image. The peak signal-to-noise ratio (PSNR) of HT-SHiLo processed images were about 14 dB higher than that of HT decoded images under different modulation depths (Figure 2(c)), and the PSNR of HiLo processed images falls between the two. As modulation depth decreased, the HT decoded images became less correlated with GT image, and the HiLo processed images also degraded at low modulation depth, while the HT-SHiLo processed images maintained high correlation with GT image (Figure 2(d)). The processing time for each algorithm was tested multiple times on GPU (RTX 3070) using MATLAB (R2023a) (Figure 2(e)). The HT-SHiLo algorithm required only an additional 3ms compared to the HT algorithm. On the other hand, the processing time for the HiLo algorithm was significantly higher for its Fourier transform, and it increased exponentially with larger image size. At 2,048 × 2,048 pixels, the HT-SHiLo algorithm was approximately 32 times faster than the HiLo algorithm.

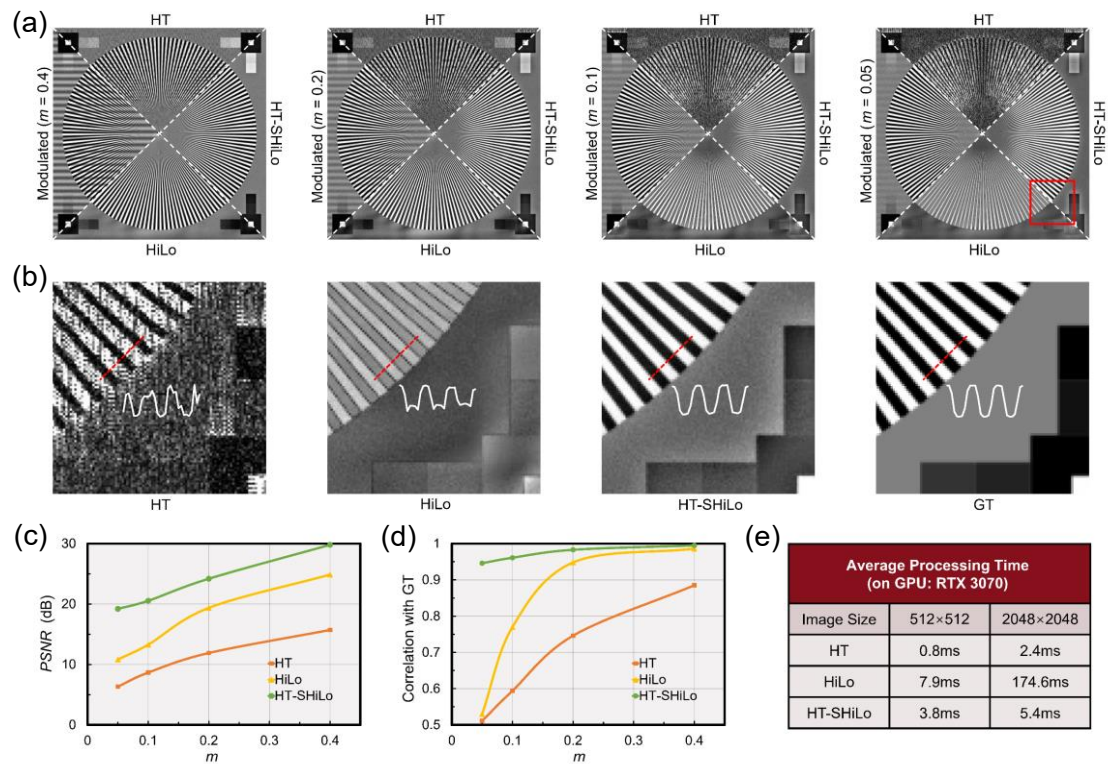

**Figure 2. Simulation of the HT decoding algorithm, HiLo algorithm and HT-SHiLo algorithm under different modulation depths.** (a) The modulated images with modulation depths $m$ = 0.4, 0.2, 0.1, 0.05 and their optical sections obtained by each algorithm. (b) Local enlargement of optical

sections obtained by each algorithm under *m* = 0.05 and the ground truth (GT) image. (c) PSNR of optical sections obtained by each algorithm under different modulation depths. (d) Pearson correlation coefficient (PCC) with GT image of optical sections obtained by each algorithm under different modulation depths. (e) Comparison of average processing time between the three algorithms.

Low modulation depth often occurs in thick tissue imaging, while fluorescence bleaching encountered during long-term recording can lead to low SNR. To address these two scenarios, we have conducted 3D imaging of thick tissues and long-term intravital imaging of cells as follows.

**Deep insight into neurons and vascular endothelial cells in mouse brain**

The neurons in animal's brain dictates a myriad of physiological processes and behaviors. Understanding how these neurons interact with each other and form a complex network in the intact brain is a fundamental question in neuroscience. To further expand the application of OS-SIM in neuroscience research, we used it to study neurons and nerves in mouse brain. An adult *Thy1-EGFP-M* mouse brain was harvested and a 2 mm thickness coronal block was sectioned and processed following the PEGASOS tissue clearing method [2][32], and excited at 488 nm. A local region with dense nerves was first imaged to test the performance of HT-SHiLo algorithm under varying imaging depth. With objective lens of 10×/0.45 NA, we got an imaging volume of 942 × 942 × 572 μm$^3$. As imaging depth increased, the noise in HT decoded 3D image became stronger (Figure 3(a)), and the structure of nerves became gradually unclear and invisible from depth of about 300 μm. In contrast, the 3D image processed by HT-SHiLo algorithm maintained high SNR and clear structure up to depth of 572 μm (Figure 3(b)). Through quantitative analysis of the SNR of optical sections (Figure 3(c)), we found that the SNR of HT decoded images decreased exponentially with imaging depth, reducing to 1/e of the peak value at depth of about 200 μm, while the SNR of HT-SHiLo processed images kept a high level till 500 μm. The mean value of SNR of HT and HT-SHiLo processed images were calculated to be 3.02 dB and 11.86 dB respectively, which means an improvement of 8.84 dB in SNR was achieved. Figure

3(d)-(g) make distinct comparisons between the optical sections processed by HT and HT-SHiLo algorithms under varying imaging depth, that HT decoded images were submerged in noise at depth of 320 and 460 μm, while HT-SHiLo processed images exhibited high contrast and SNR at all depths.

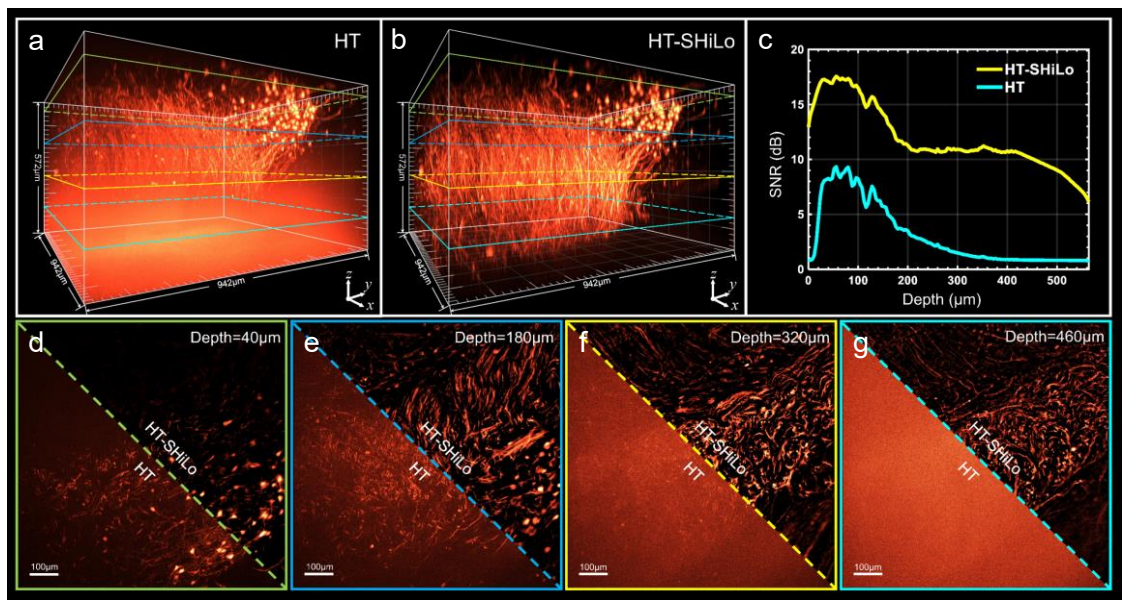

**Figure 3. Comparison between HT and HT-SHiLo algorithms on imaging of nerves in mouse brain under varying imaging depth.** (a) Lateral view of HT decoded 3D image. (b) Lateral view of HT-SHiLo processed 3D image. (c) SNR of optical sections under varying imaging depth. (d)-(g) Comparison of optical sections under imaging depths of 40, 180, 320 and 460 μm.

To evaluate large tile imaging, we acquired a large block of mouse brain from *Thy1-EGFP-M* sample at the hippocampus region, representing the largest coronal dimension of the mouse brain. The sample was cleared using the PEGASOS tissue clearing method. We first scanned at 5 × 3 FOVs with 4×/0.2 NA objective lens to acquire 3D image of the whole brain slice which is 9.4 × 6.2 × 1.3 mm$^3$ in size (Figure 4(a), Video S1). The *Thy1-EGFP-M* mouse model specifically labels excitatory neurons, including both their somas and long axonal projections. Typically, neuron somas are around 10 μm in size, while axons measure approximately 0.5 μm. In the low magnification image, all neuron somas are clearly visible (Figure 4(a)). To capture more detailed features, a 20×/0.75 NA objective lens was used to obtain higher resolution images (Figure 4(b)-(d)). All somas and axons in the imaging blocks were clearly identified with comparable resolution, and depth-coded colors help distinguish the height distribution of each axon. Dendritic spines are small protrusions on the dendrites of neurons that play a crucial role in synaptic plasticity and the formation of neural networks. Notably, dendritic spines of micrometer size can be observed densely distributed throughout the space (Figure 4(e)-(g)), confirming the high resolution of our system.

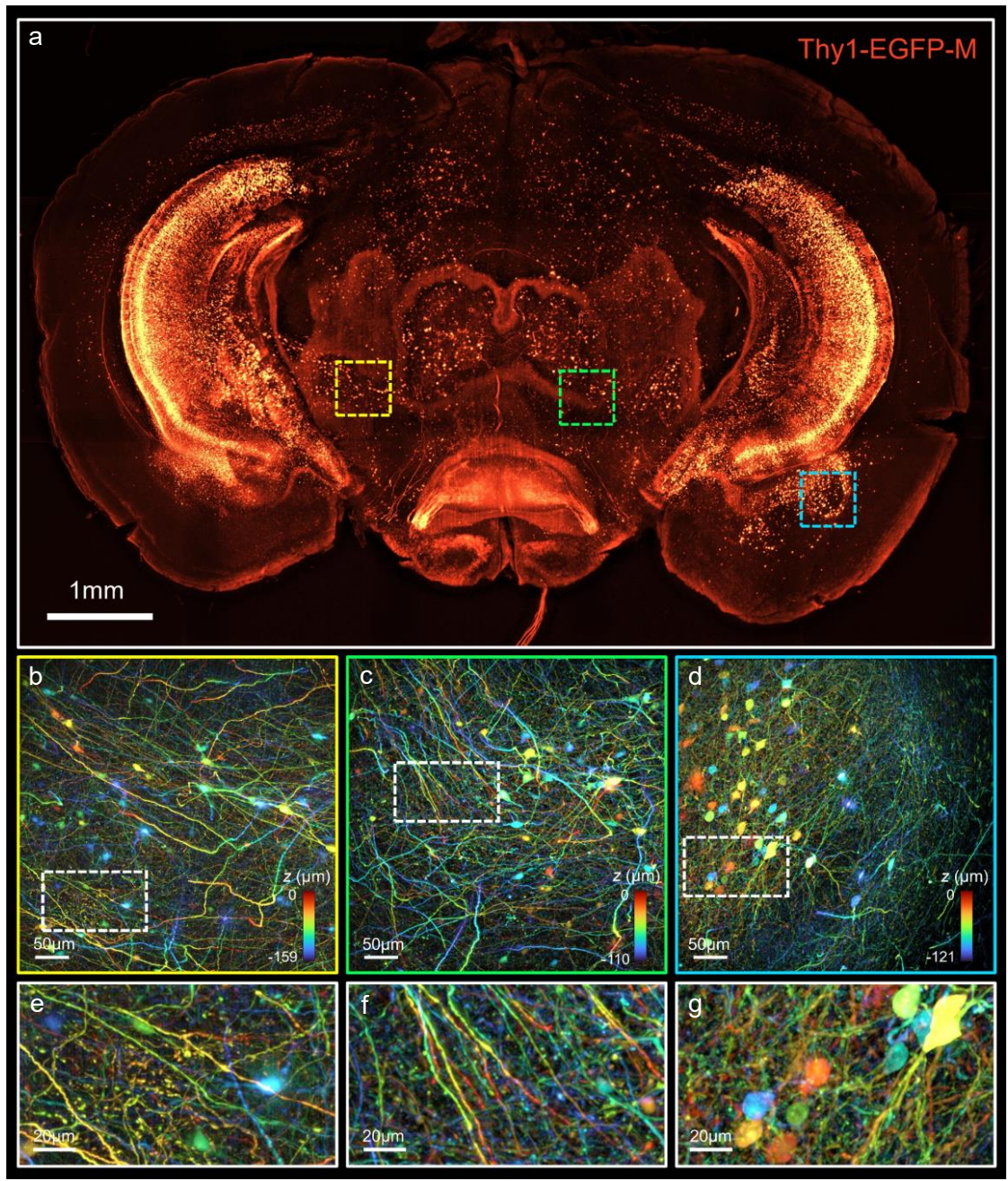

**Figure 4. 3D images of nerves and neurons in a mouse brain slice.** (a) Maximum intensity projection (MIP) image of the whole brain slice. (b)-(d) Height-coded MIP images at corresponding regions in (a). (e)-(g) Local enlargement of (b)-(d). (See also Supplementary Video S1)

We furtherly imaged the vasculature within a mouse brain. We acquired a large block of mouse brain from *Tie2-Cre; Ai14; H2B-GFP$^{flox}$* mouse. In this model, all vasculature nuclei were specifically labelled with GFP fluorescence while their cytoplasm was labeled with tdTomato fluorescence, respectively excited at 488 nm and 561 nm. The sample was cleared with the PEGASOS method. The whole brain slice was scanned at 5 × 3 FOVs with 4×/0.2 NA objective lens, resulting in a 3D image size of 8.6 × 5.6 × 2.4 mm$^3$ (Figure 5(a), Video S2). Low magnification image revealed the overall organization of brain vasculature throughout the entire coronal section. Detailed images were captured using objective lens of 10×/0.45 NA (Figure 5(b)-(d)). In all indicated regions, nuclei with dimension of ~5μm and cytoplasm with dimension of ~10μm were both clearly visualized. Achieving an imaging depth of approximately 800 μm (10×/0.45 NA objective) in such a densely packed sample demonstrates our system's deep 3D imaging capability.

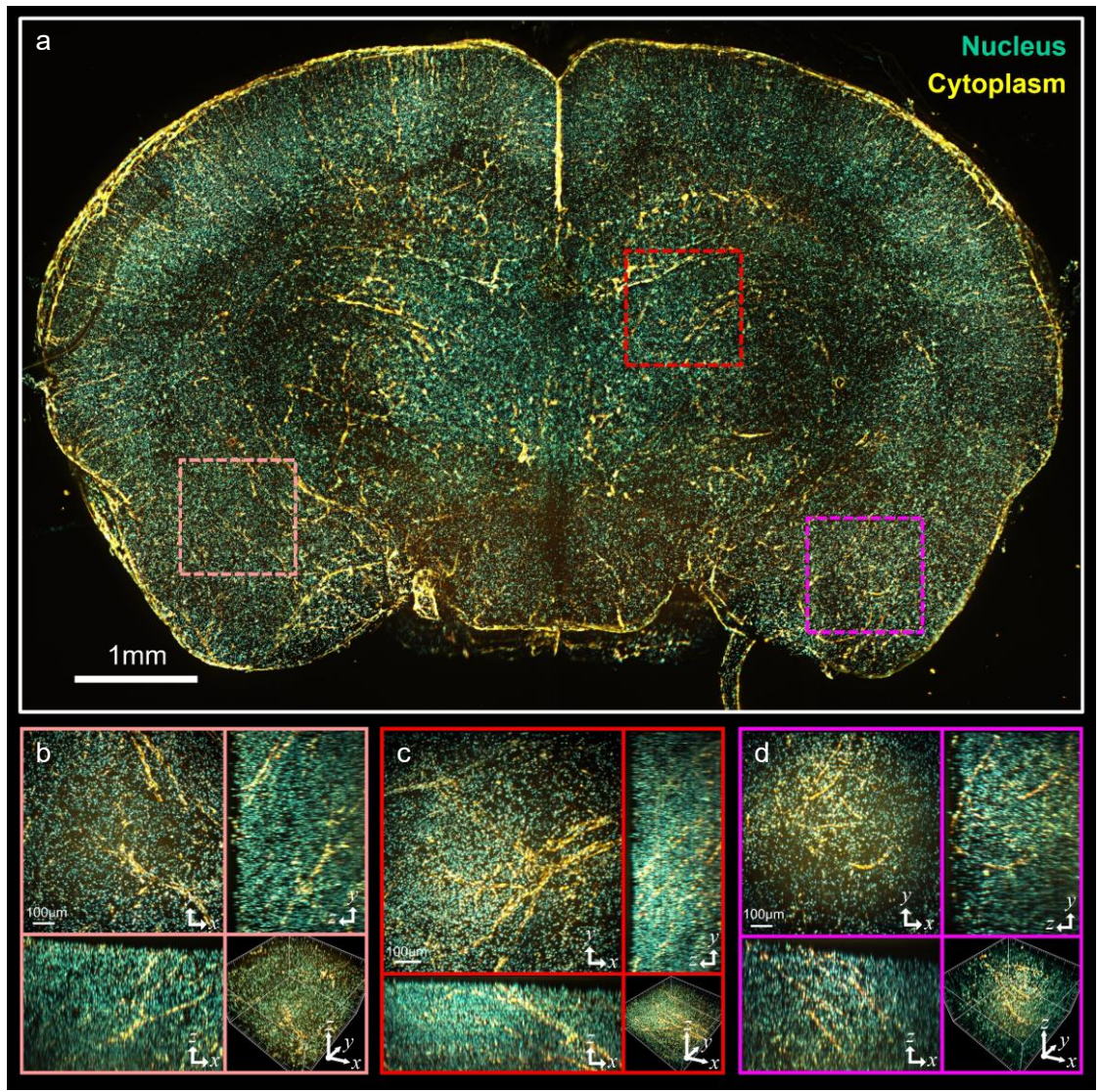

**Figure 5. 3D image result of nucleus (green) and cytoplasm (yellow) of vascular endothelial cells in a mouse brain slice.** (a) MIP image of the whole brain slice. (b)-(d) 3D images of local regions in (a) with their MIP images along each axis. (See also Supplementary Video S2)

**3D imaging of the clock neurons in intact Drosophila brains**

Drosophila are invaluable for studying the underpinnings of behavior and neurological disorders due to their genetic simplicity and conservation of key neural processes with humans. We applied OS-SIM to study of Drosophila clock neurons, a relatively small network regulating the locomotor activity and sleep of these animals. A pan-clock neuron driver, Clk21-Gal4, was crossed with UAS-EGFP to label the clock neurons with excitation at 488nm. The brains were dissected and fixed before undergoing the standard immunostaining. NC82 staining was used to label synaptic neuropil regions in the brain with excitation at 635 nm. With 40×/1.3 NA oil immersion objective lens, the brain was scanned at 3 × 2 FOVs, resulting in a FOV size of 584 × 317 μm$^2$ and depth of 123 μm. OS-SIM offered a wide-field view of clock network in the intact brain and, at the same time, its descent resolution revealed the precise projection patterns of clock neurons (Figure 6(a), Video S3). More importantly, there was much less signal attenuation between the lateral and dorsal clock neurons. The height map enabled a better understanding of the connections within the clock neuron network, as well as its interactions with other neuron circuits (Figure 6(b-d)). Combined with MultiColor FlpOut (MCFO) and circuit trancing techniques such as trans-Tango, we envision OS-SIM becoming a powerful toolkit to study neuronal network in fly brains.

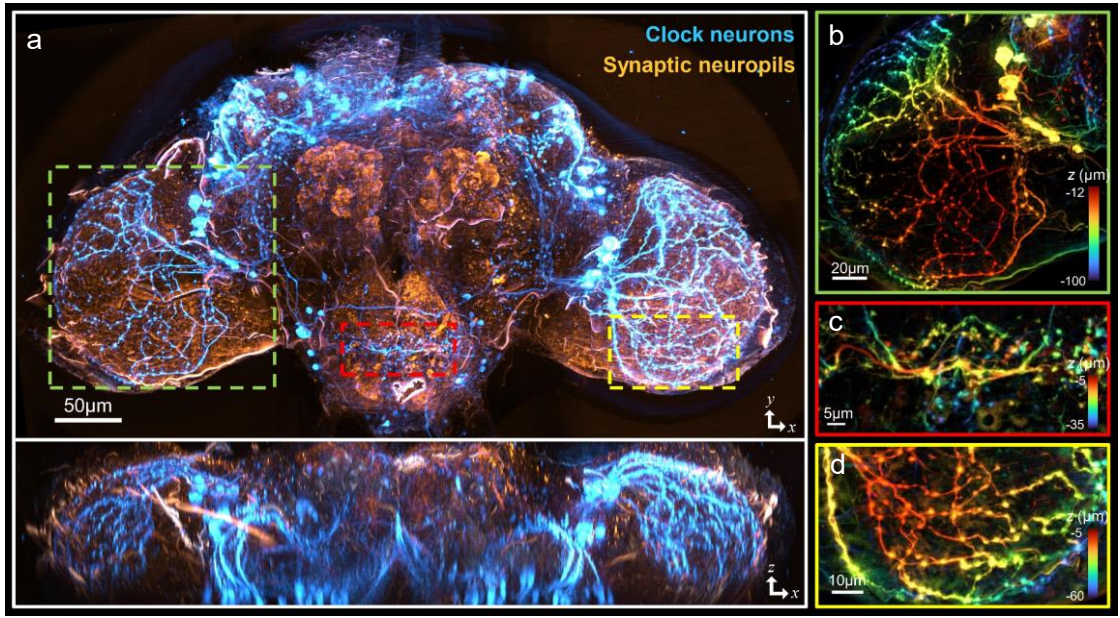

**Figure 6. 3D imaging result of clock neurons (blue) and synaptic neuropils (yellow) in a Drosophila brain.** (a) MIP images of the whole brain along z and y axes. (b-d) 3D images of clock neurons in local regions. (See also Supplementary Video S3)

**Tomographic imaging of human colonic organoids**

Organoids are tissue analogues with certain spatial structure formed by 3D culture of tissue samples or pluripotent stem cells in vitro, which can simulate real organs in structures as well as functions. Colonic organoids can reproduce the 3D features of colon tissue, especially the specific crypt-villi structure of colon epitheliums. The PFA-fixed and stained human colonic organoid sample was bestowed by Genome Institute, The First Affiliated Hospital of Xi'an Jiaotong University. The sample was stained with Ki67 to indicate cell growth, and with E-cadherin to indicate cell adherens junction, respectively excited at 385 nm and 488 nm. The organoid was scanned at 3 × 3 FOVs with 60×/1.49 NA oil immersion objective lens, which is $350 \times 364 \times 53$ μm$^3$ in size. The adherens junction within cells is scanned clearly, which are expressed only on the periphery of colonic organoid (Figure 7, Video S4). The scanned result indicates the well differentiated zones, where the colonic epithelium forms (Figure 7(a), red arrows), and the less differentiated zones (Figure 7(a), green arrows), as the E-cadherin is one of the markers to reflect enterocytes. The bulging area of Figure 7(b) showed the growing villi (purple arrow) and crypt (cyan arrow). The colonic cavity can also be scanned with high spatial resolution, that is, the cavity morphology changes with different depth (Figure 7(c), white arrows), indicating the variational structure within the organoid.

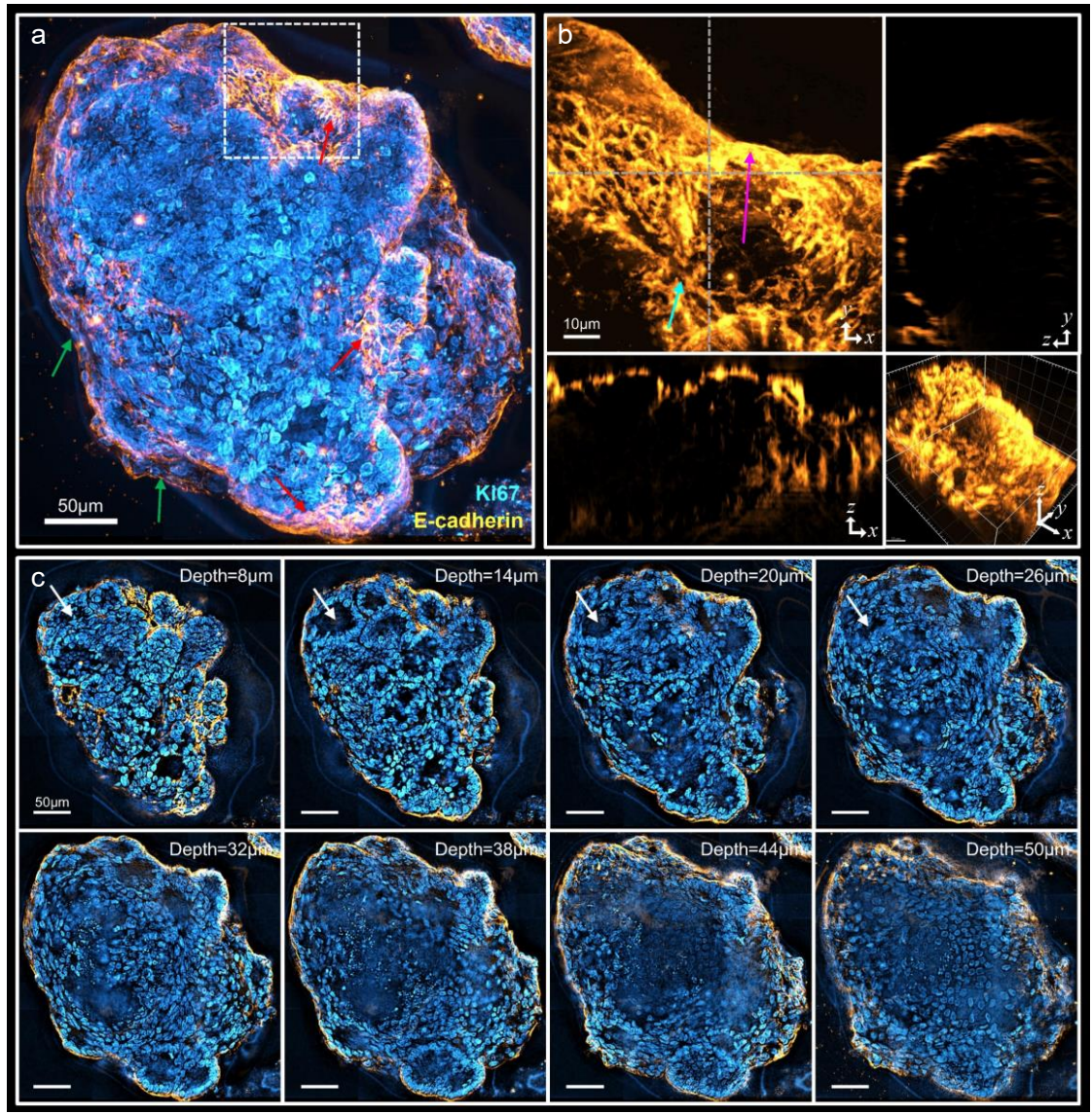

**Figure 7. Tomographic imaging of Ki67 (blue) and E-cadherin (yellow) of a human colonic organoid.** (a) MIP image of the organoid. (b) MIP and 3D image of E-cadherin in the local region in (a), with the *xoz* and *yoz* sections along gray dashed lines. (c) Optical sections of the organoid at various imaging depths. (See also Supplementary Video S4)

## High SNR imaging of mitochondria in live cells

In live-cell imaging, high light power can lead to photobleaching and phototoxicity, which may hinder the observation of cellular activities over an extended period. To evaluate the performance of HT-SHiLo algorithm in noise suppression with reduced light power, we first applied OS-SIM on imaging of cell nucleus and mitochondria in BSC-1 cells. Cell nucleus stained with Hoechst 33342 and mitochondria stained with

Alexa Fluor™ 555 were respectively excited at 385 nm and 561 nm. The used light power density was about 1 W/cm$^2$ at the focal plane, and the same in subsequent live-cell experiments. An Oil immersion objective lens of 100×/1.49 NA was employed with a field of view (FOV) size of 93 × 93 μm$^2$, and a scan range of 11 μm along z-axis was adopted. The maximum intensity projection (MIP) image of the BSC-1 cell revealed distinct contrast and background subtraction (Figure 8(a), Video S5), with the outer membrane of the mitochondria clearly visible. The HT-SHiLo algorithm exhibited excellent noise suppression compared with HT algorithm, and also removed the residue fringes (Figure 8(b, c)). The SNR of MIP images were calculated to be 20.53 dB and 9.91 dB respectively for HT-SHiLo and HT algorithm, which means a 10.62 dB improvement in SNR was achieved. With high spatial resolution achieved by OS-SIM, the 3D morphology of mitochondria could be precisely observed (Figure 8(d, e)).

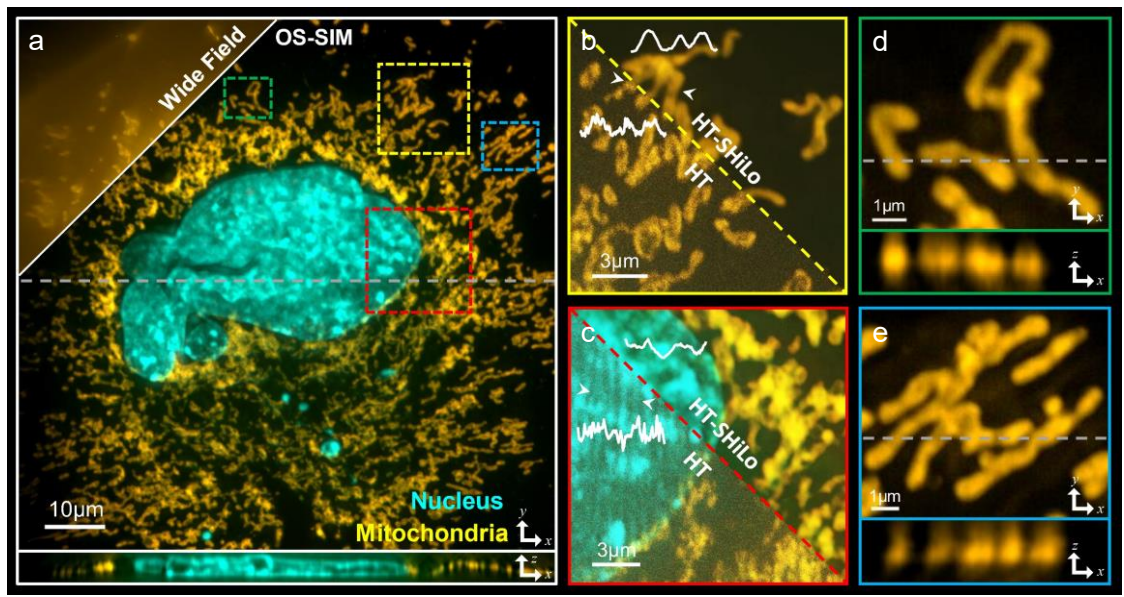

**Figure 8. High SNR imaging of cell nucleus (cyan) and mitochondria (yellow) in BSC-1 cells.** (a) MIP image of the BSC-1 cells with the *xoz* section along the gray dashed line. (b, c) Comparisons between the results of HT and HT-SHiLo algorithms. (d, e) MIP images of mitochondria with *xoz* sections along the gray dashed lines. (See also Supplementary Video S5)

Mitochondrial dynamics, encompassing fusion, fission, and transportation, constitute essential regulatory processes for organelle quality control [33]. The precise regulation of these dynamic processes is fundamental to cellular homeostasis and has profound implications for various physiological and pathological states [34][35]. While mitochondrial dynamics inherently occur in three dimensional spaces, the rapid nature of fusion and fission events limits the usage of 3D methods for studying the kinetics of mitochondrial dynamics in living cells. The superior spatiotemporal resolution and three-dimensional imaging capabilities of OS-SIM provide an unprecedented opportunity to address the long-standing challenge of quantifying mitochondrial dynamics in three dimensions. We exploited this advanced imaging technique to capture and analyze comprehensive mitochondrial dynamics in living COS-7 cells. Stained with PK Mito Orange and excited at 561 nm, the mitochondria were imaged using 100×/1.49 NA oil immersion objective lens and a z-scan range of 3.2 μm to cover their stacked thickness (Figure 9(a), Video S6). The dynamics of mitochondria were recorded for 5 minutes with 10-seconds intervals.

  Taking advantage of the rapid three-dimensional imaging capabilities of OS-SIM, we obtained unprecedented spatiotemporal resolution of mitochondrial dynamics across multiple Z-planes. Our observations revealed detailed mitochondrial behaviors, including elongation and transport along microtubules, followed by subsequent network dissociation (Figure 9(b)). Notably, post-fission mitochondria exhibited increased structural flexibility and enhanced mobility within the cytoplasm. The superior Z-axis resolution enabled direct visualization of membrane tubule formation preceding mitochondrial fusion events, occurring both within the same focal plane and across different Z levels (Figure 9(c, d)). These observations provide direct evidence supporting previously proposed fusion mechanisms derived from structural biology studies [36].

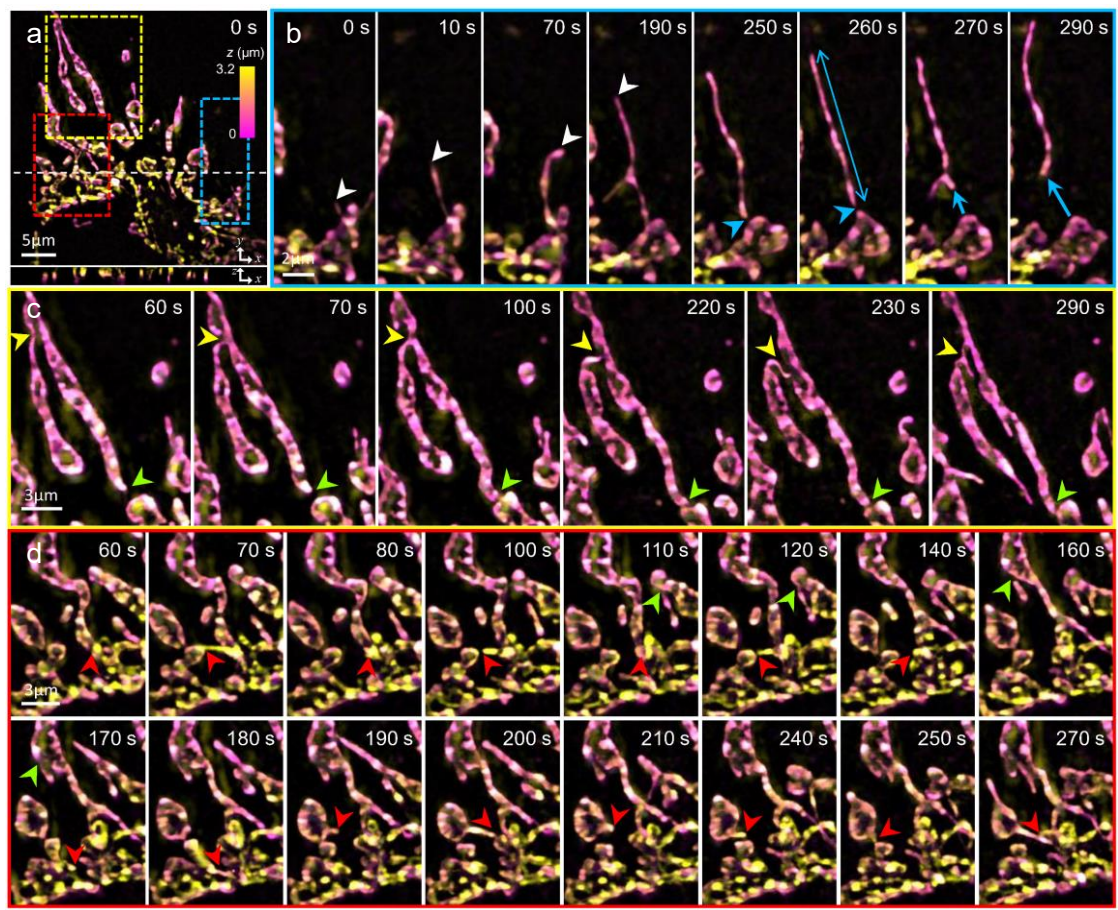

**Figure 9. Dynamics of mitochondria in COS-7 cells.** (a) Height-coded MIP image of the mitochondria at the initial moment with the *xoz* section along the white dashed line. (b) Mitochondrial fission in the blue box in (a). (c) Mitochondrial fusion in the yellow box in (a). (d) Mitochondrial kiss and fusion in the red box in (a). (See also Supplementary Video S6)

**Long-term intravital imaging of migrasome dynamics**

Migrasomes are cell organelles produced during cell migration. At the trailing edge of a migrating cell, retraction fibers form, and migrasomes develop at their cross-points or tips into vesicular structures measuring 0.5 to 3 μm in diameter [37][38]. Migrasomes act as carriers for a diverse range of proteins, nucleic acids, and lipids, mediating lateral transfer of materials and information, and establishing micro-environments that play crucial roles in embryonic development, angiogenesis, blood coagulation, and immune responses [39][40][41]. Tetraspanin4, a four-transmembrane protein, is a recognized marker of migrasomes and plays a critical role in their formation [42].

To further elucidate the mechanism and function of migrasomes, we employed the OS-SIM to visualize them in high-quality, three-dimensional live-cell images, capturing detailed information. Using a 100×/1.49 NA oil immersion objective lens and excitation at 488 nm, we acquired images with a z-scan range of 3.8 μm, recording the dynamic processes of migrating cells over 80 minutes with 1-minute intervals (Figure 10a, Video S7). The time-lapse imaging showed that the cell left behind retraction fibers at its tail during migration (Figure 10b). The migrasome grew from 0.5 μm to 1.2 μm in x-y diameter and from 0.8 μm to 2 μm in height during 80 minutes, which also indicated that the migrasome was ellipsoidal with an approximate aspect ratio of 1:1.6 (Figure 10c). We also captured the fragmentation of retraction fibers which tended to break at the connection points with migrasomes (Figure 10d). However, migrasomes remained attached at the local site and retained their complete structure even after detaching from the retraction fibers. These observations suggest that migrasome membranes may seal before retraction fiber breakage, potentially endowing them with properties characteristic of extracellular vesicles at later stages.

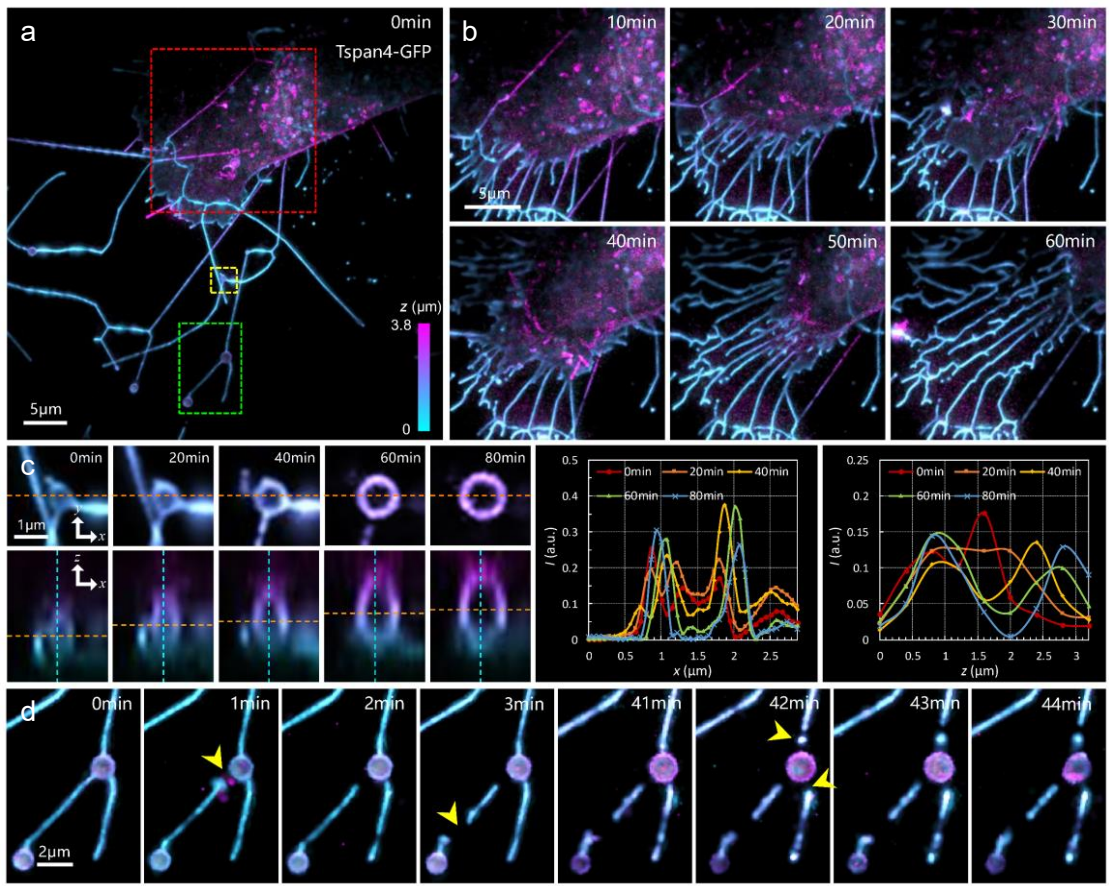

**Figure 10. Dynamic imaging of a migrating Tspan4-GFP NRK cell and its migrasomes.** (a) Height-coded MIP image of the migrating cell at the initial moment. (b) The process of cell migration leaving behind retraction fibers corresponding to the area enclosed by red box in (a). (c) The growth process of the migrasome located within the yellow box in (a). Section images on *xoy* plane and *xoz* plane along the orange line were respectively shown, and the intensity variances along the orange line and blue line were measured. (d) The fragmentation process of the migrasome located within the green box in (a). (See also Supplementary Video S7)

## Discussion

For 3D imaging of biological tissues and structures, OS-SIM stands out for its high speed, low phototoxicity, high spatial resolution and broad applicability. However, the traditional RMS decoding method leads to two problems: First, three raw images are required for reconstruction of one optical section, which not only implies a relatively low imaging efficiency but also brings more phototoxicity and photobleaching to the sample; Second, the SNR of optical section relies on the modulation depth of fringe that faces severe reduction when imaging thick tissues. HT decoding method reduces the required raw images to two, but the high frequency part of HT decoded optical section undergoes a deviation, and it faces the same problem of low SNR. Applying the HiLo processing can effectively improve the SNR of optical section to the same level of wide field image, but the traditional HiLo algorithm involves the time-consuming Fourier transform. Therefore, we proposed the HT-SHiLo algorithm which performs the HiLo operation entirely in space domain and thus with a 32 times promotion in processing speed. Meanwhile, the inaccurate high frequency part of HT decoded optical section can be revised. In general, HT-SHiLo algorithm provides an elegant solution for fast and high SNR optical sectioning via SIM.

With our OS-SIM system, we successfully captured 3D images of mouse brains, Drosophila clock neurons, organoids, mitochondria and migrasome. The results demonstrate high SNR image production using HT-SHiLo algorithm compared to directly HT decoding with an improvement of about 10 dB in SNR. Meanwhile, the imaging depth can be doubled thanks to the information fidelity of HT-SHiLo algorithm under noise submergence, which may expand the application of OS-SIM in thick tissues imaging. Incorporating adaptive optics (AO) into SIM can also expand imaging depth and correct aberration [17][43], but it leads to sophisticated imaging system and high cost.

The temporal resolution of current system is mainly limited by the scanning speed of z-axis stage and the excitation power of fluorescent probe. With exposure time of 10 ms and stage translation time of 10 ms, the highest volumetric imaging frame rate is about 3.3Hz when scanning 10 layers (for cells sample with thickness of 4 μm and z-step of 0.4 μm). It can be further improved through replacing the translation stage with a faster focal plane changing solution, like TAGLENS which can zoom in sub-millisecond. Brighter fluorescent probes can also help reduce the exposure time. At the shortest exposure time and the fastest axial scanning speed using TAGLENS, the optical sectioning speed can reach up to nearly half of the camera frame rate (400fps @ 512 × 512 pixels). In this case, scanning cells sample could achieve a volume frame rate up to nearly 20 Hz, which allows in vivo 3D imaging of fast biological processes.

The current system's spatial resolution has already reached the optical diffraction limit. Further improvement in resolution can be achieved by applying image reprocessing like sparse deconvolution [44], multi-resolution analysis (MRA) deconvolution [45], deblurring by pixel reassignment (DPR) [46], zero-shot deconvolution networks (ZS-DeconvNet) [47] and so on. Although 3D-SIM can physically provide higher resolution 3D imaging [48][49][50], it requires 15 raw images for reconstruction of one layer, which is not suitable for fast dynamic 3D imaging and can record fewer time points compared with OS-SIM. Therefore, OS-SIM is a quite promising technique for research of the evolution and functional processes of cellular organelles or tissues in 3D morphology.

## Materials and Methods

**OS-SIM system**

The OS-SIM system for 3D fluorescence imaging is illustrated in Figure 11. The illumination light is emitted by LEDs (X-Cite Turbo, Excelitas Inc., Canada) with switchable wavelengths of 385 nm, 488 nm, 561 nm and 635 nm, and passes through the excitation filter. Entering the total internal reflection (TIR) prism, the light is reflected onto the DMD (V7000, ViALUX GmbH, Germany) to be modulated into a binary grating. The modulated light field then passes through the achromatic tube lens ($f$ = 300 mm) and the dichroic mirror, and is projected onto the focal plane of the objective lens (4×/0.2 NA, TL4X-SAP, Thorlabs Inc., USA; 10×/0.45 NA, S Fluor, Nikon Inc., Japan; 20×/0.75 NA, Plan Fluor, Nikon Inc., Japan; 40×/1.3 NA, Plan Fluor, Nikon Inc., Japan; 60×/1.49 NA, Apo TIRF, Nikon Inc., Japan; 100×/1.49 NA, Apo TIRF, Nikon Inc., Japan). The illumination pattern on sample becomes sinusoidal distribution naturally due to the limited bandwidth of objective lens. The sample is mounted on an XYZ motorized translation stage (M-112.1DG1, 25 mm travel range, Physik Instrumente GmbH and Co. KG, Germany) that volume data can be obtained by axially moving the stage. Meanwhile, moving in landscape orientation with image stitching applied enables the extension of the FOV. For fast imaging of dynamic process, we use the piezo stage (P-726 PIFOC, 100 μm travel range, Physik Instrumente GmbH and Co. KG, Germany) to axially scan the objective. A sCMOS camera (ORCA-Flash4.0 V3, 100 fps @ 2,048 × 2,048 pixels, 16 bits, Hamamatsu, Japan) is employed to capture the 2D image with an emission filter blocking the excitation light. A custom software programmed in C++ is utilized to control DMD patterns generation, stage movement and image record. A 100 nm fluorescent bead (505 nm excitation, 515 nm emission) is used to measure the point spread function (PSF) of the OS-SIM system (Figure 11c). OS-SIM removes the background in wide-field image, and the full width at half maximum (FWHM) of the bead's PSF is measured to be 280 nm in $x$ and 510 nm in $z$ under 60×/1.49 NA oil immersion objective.

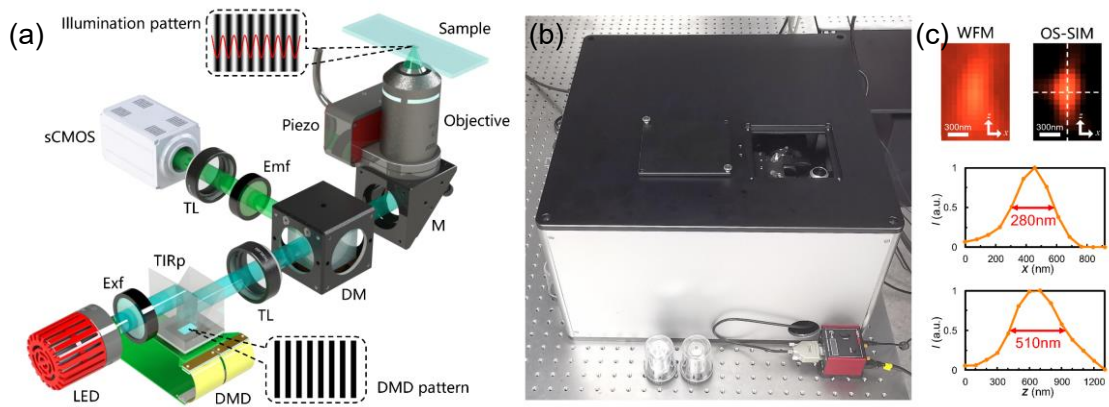

**Figure 11. The OS-SIM system.** (a) Schematic diagram of the light path. DM, dichroic mirror; DMD, digital micromirror devices; Emf, emission filter; Exf, excitation filter; LED, light emitting diode; M, mirror; sCMOS, scientific complementary metal-oxide semiconductor; TIRp, total internal reflection prism; TL, tube lens. (b) Photograph of the custom-built instrument. (c) XZ views of a 100 nm fluorescent bead under 60×/1.49 NA oil immersion objective, with its intensity obtained by OS-SIM along $x$ and $z$ directions measured.

**PEGASOS tissue clearing method for preparing mouse brain samples**

All samples were collected and processed using PEGASOS tissue clearing kit (Leads Bio-Tech, http://www.leads-tech.com Catalogue number: PSK100N). A Mice were transcardially perfused with PBS followed by 4% paraformaldehyde (PFA). The brains were then dissected and fixed in 4% PFA overnight at 4 degrees. They were then processed following the kit protocol. Finally, samples were immersed in BB-PEG medium (solution 7 in the kit) for at least one day for clearing. Samples were immersed in the clearing medium for imaging.

**Drosophila brain preparation**

Flies were raised on standard cornmeal medium with yeast under LD (12 h light/12 h dark) conditions at room temperature. Three to five days old flies were entrained in LD conditions for 3 days before being fixed with 4% (vol/vol) paraformaldehyde with 0.5% Triton X-100 for 2 hr and 40 min at room temperature. Brains were dissected and then washed twice (10 min) in 0.5% PBST buffer before being blocked overnight in 10% Normal Goat Serum (NGS; Absin) at 4℃. The brains were then incubated in chicken anti-GFP antibody at a 1:1000, a mouse anti-NC82 antibody at 1:30 for overnight, then the brains then washed twice (10 min) in 0.5% PBST buffer. The corresponding secondary antibodies were added and incubated overnight. Brains were mounted in KIT04 Mounting Medium Sunbloss (HUAXIA Imaging).

**Cell culture and staining of cells**

The Biologics Standards-Cercopithecus-1 (BSC-1) cell line was purchased from Pricella Life Technology Co. LTD. COS-7 cell line was purchased from Applied Biological Materials Inc. Both cells were cultured in DMEM (Invitrogen, #11965-118) supplemented with 10% fetal bovine serum (FBS) (Gibco, #16010-159). To prevent bacterial contamination, 100 μg/ml penicillin and streptomycin (Invitrogen, #15140122) were added in the DMEM medium. Cells were grown under standard cell culture conditions (5% $CO_2$, humidified atmosphere at 37°C).

For the immunofluorescence staining of BSC-1 cells, cells were seeded in a 35 mm cell dish with a #1.5 glass coverslip (SunBloss™, STGBD-035-1). To increase cell adhesion, we pre-treated the glass-bottom dishes with fibronectin (SunBloss™, HXAR-01) for 1 h at 37°C. When the cells grown to a confluence of around 60%, they were fixed with 37°C pre-warmed fixation buffer for 10 min, containing 4% paraformaldehyde and 0.1% glutaraldehyde (SunBloss™, HXKx01) in PBS. Then the cells were washed three times with PBS. To quench the background signals, we incubated the cells with 2 ml of 0.1% $NaBH_4$ solution (SunBloss™, HXIK023) in PBS for 7 min, and optionally shook the dish on a shaker (<1Hz). The cells were washed three times with 2 ml PBS and then incubated for 30 min in PBS containing 5% BSA and 0.5% Triton X-100 (SunBloss™, HXKx02) at 37°C. All antibodies were diluted in the 5% BSA+ 0.5% Triton X-100 solution described above. Next, we incubate the cells for 40 min with the appropriate dilution of primary antibody: Tom20 (abclonal, A19403) at 25°C. After primary antibodies incubation, the cells were washed 5 min with 2 ml PBS for three times. Secondary antibody and dye were incubated for 60 min with the appropriate dilutions of them: 1:100 Goat anti-Rabbit IgG (H+L) Cross-Adsorbed Secondary Antibody, Alexa Fluor™ 555 (Thermofisher, A-21428) and 50 umol/ml Hoechst 33342 (Thermofisher, 62249) at 25°C. After washed 3 times with PBS, cells were fixed with post-fixation buffer (SunBloss™, HXIE-2) for 10 min.

For the mitochondria labeling of COS-7 cell line, cells were seeded in a 35 mm cell dish with a glass bottom and incubated overnight. When the cells grown to a confluence of around 60%, cells were washed with PBS for three times and fresh culture medium containing PK Mito Orange (Nanjing Genvivo Biotech Co., Ltd.) at a concentration of around 200 nM was added into the cell dish. After additional incubation for 20 min, the COS-7 cells were washed with PBS for three times and fresh culture medium was added. The PK Mito Orange stained cells were placed on the microscope stage for imaging.

**Sample preparation for migrasome observation**

The imaging chamber was coated with fibronectin at a concentration of 10μg/ml, and the coating process took place for a minimum of 30 minutes at 37°C, but not exceeding 2 hours. Following the coating, Tspan4-GFP stable over-expressed NRK (normal rat kidney) cells were seeded onto the pre-coated glass bottom imaging chamber. The cells were cultured in DMEM (Dulbecco's Modified Eagle Medium) supplemented with 10% FBS (Fetal Bovine Serum) and 1× Penicillin-Streptomycin Solution. The culture was maintained at 37°C with 5% $CO_2$ for approximately 15 hours. Subsequently, the sample was ready for further imaging analysis.

**Computation of PSNR and SNR**

The PSNR used in simulation where GT image is known is computed as follow:

$$PSNR(\text{dB}) = 10\log_{10}\frac{1}{(I-I_{GT})^2} \quad (10)$$

The SNR used in experiment where GT image is unknown is computed as follow:

$$SNR(\text{dB}) = 10\log_{10}\frac{std[I]^2}{stdfilt[I]^2} \quad (11)$$

where *std*[ ] calculates the standard deviation of image, while *stdfilt*[ ] performs standard deviation filtering where the value of each output pixel is the standard deviation of the 3-by-3 neighborhood around the corresponding input pixel.

**Computation of Pearson correlation coefficient**

Pearson correlation coefficient (PCC) is used to evaluate the similarity with GT image:

$$PCC[I, I_{GT}] = \frac{Cov[I, I_{GT}]}{std[I] \cdot std[I_{GT}]} \tag{12}$$

where $Cov[I, I_{GT}]$ represents the covariance of $I$ and $I_{GT}$.

**Author Contributions**

M.L. and M.W. conceived the project. M.W., M.S., J.Y. and C.L. realized the setup of apparatus and conducted the experiment. M.W. wrote the code and performed the data analysis. M.W. and M.L. wrote the paper. M.L. supervised the research project. All authors participated in reading and editing of the final paper.

**Funding**

This work was supported by the Natural Science Foundation of China (NSFC) (Nos. 62135003, 62205267, 62205265); National Key Research and Development Program of China (2022YFF0712500); The Innovation Capability Support Program of Shaanxi (No. 2021TD- 57); Natural Science Basic Research Program of Shaanxi (Nos 2022JZ-34, 2022JM-321); the Shaanxi Fundamental Science Research Project for Mathematics and Physics (22JSY028) and the Qinchuangyuan Talent Program (QCYRCXM-2022-312)